\documentclass[
 reprint,
 amsmath,amssymb,
 aps,
]{revtex4-1}

\usepackage{graphicx}
\usepackage{dcolumn}
\usepackage{bm}
\usepackage{xcolor}
\usepackage{soul}

\begin{document}

\preprint{APS/123-QED}

\title{Electromagnetic and Acoustic Fano Interference in Surface Acoustic Wave Resonators}

\author{P.K.~Rath}
 \email{rathpra1@msu.edu}
\affiliation{Department of Physics and Astronomy,
Michigan State University, East Lansing MI 48824 USA
}

\author{J.M.~Kitzman}
\altaffiliation[Present address: ]{Quantinuum, Brooklyn Park 55422, USA}
\affiliation{Department of Physics and Astronomy,
Michigan State University, East Lansing MI 48824 USA
}

\author{M.~Mesbah}
\affiliation{Department of Physics and Astronomy,
Michigan State University, East Lansing MI 48824 USA
}

\author{C.~Undershute}
\affiliation{Department of Physics and Astronomy,
Michigan State University, East Lansing MI 48824 USA
}

\author{J.~Pollanen}
\email{pollanen@msu.edu}
\affiliation{Department of Physics and Astronomy,
Michigan State University, East Lansing MI 48824 USA
}

\date{\today}

\begin{abstract}
Surface acoustic wave-based resonators are sensitive probes of condensed matter systems, as well as surface-selective sensors for chemistry and biology. Surface acoustic wave devices have also been integrated into hybrid quantum systems with qubit platforms for applications in quantum information processing and sensing. The sensitivity of piezoelectric surface wave-based resonators to investigate these various systems can be enhanced by optimizing the device architecture and electrical measurement techniques. Alternatively, tailoring the spectral symmetry, arising from interference effects, offers a promising route to further improve sensitivity. In this work, we demonstrate the simultaneous introduction of both electromagnetic and acoustic Fano interference to shape the spectral response of GHz-frequency surface acoustic wave resonators. By systematically modifying the acoustic reflectivity of the resonators, we are able to isolate and analyze each interference mechanism independently. The broad range of temperature operation, from ambient to cryogenic temperatures highlights the potential for both classical and quantum sensing applications.
\end{abstract}

\maketitle

\section{Introduction }

\begin{figure*}[ht]
    \centering
    \includegraphics[width=1\linewidth]{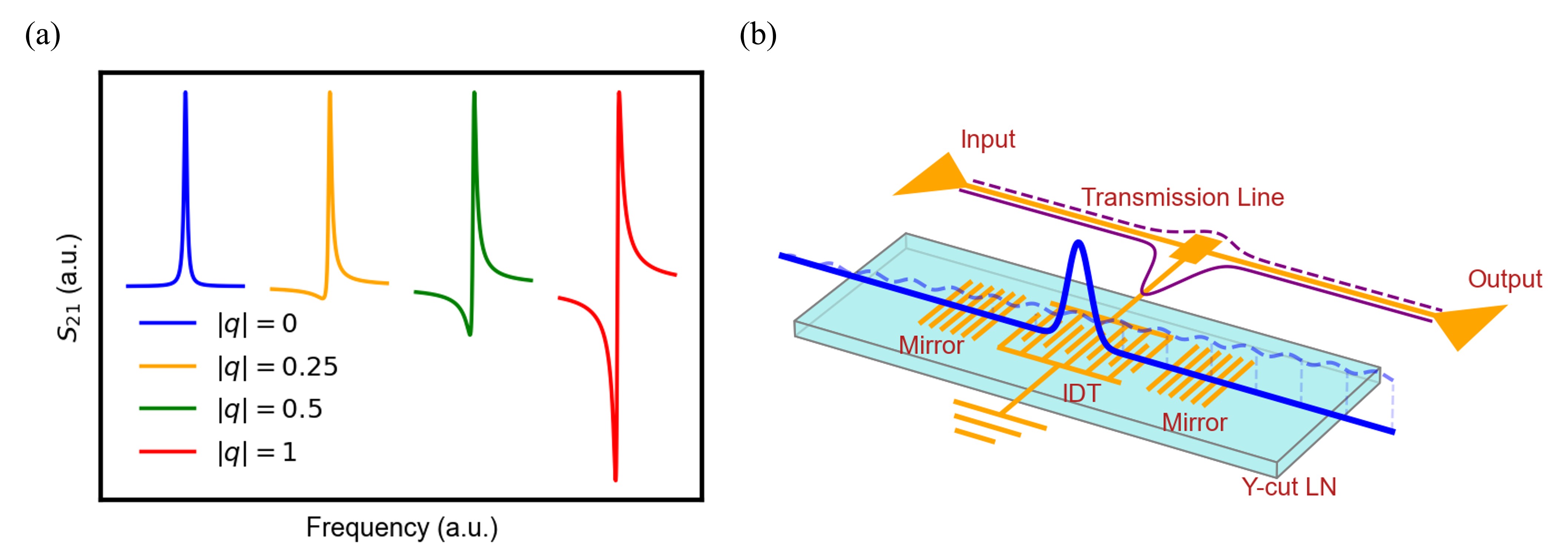}
    \caption{~(a) Schematic of the spectral line shape of a generic resonator for different values of the Fano parameter ($q$). The case when $q$ equals 0 corresponds to the absence of Fano interference and results in a Lorentzian line shape (see Eq.~1). Increasing Fano interference, i.e., increasing $|q|$, results in an increasing asymmetry of the spectral line.~(b) Schematic of the notch-type transmission measurement configuration of a surface acoustic wave device on Y-cut lithium niobate. The input microwave signal (solid purple) is injected via a transmission line waveguide and piezoelectrically transduced into resonant surface acoustic wave modes (solid blue) and a background acoustic continuum (dashed blue). The interference between these resonant and continuum signals leads to acoustic Fano interference. The resultant signal is transduced back into a microwave signal (solid purple), which can also interfere with a directly transmitted electromagnetic background through the waveguide (dashed purple).}
    \end{figure*}

Surface acoustic waves (SAWs) are mechanical oscillations that travel along the surface of a crystal, localized to approximately a wavelength above and below the surface. 
These waves can be generated on the surface of piezoelectric crystals via time-varying electric fields applied to metallic surface transducers that convert electrical signals into mechanical waves. Owing to their strain and piezoelectric coupling, SAWs provide a versatile platform for controlling and probing many condensed-matter systems. SAW techniques have been widely used to probe frequency-dependent conductivity in low-dimensional many-body quantum matter~\cite{Willett1993,Drichko2016,Pollanen2016,Friess2017,Wang2025} and to create tunable acoustic lattices for manipulating collective states~\cite{cerda2017quantum, Schuetz2017}. They have enabled coherent transport of individual charges ~\cite{hermelin2011electrons, mcneil2011demand}, spins~\cite{bertrand2016fast}, and single photon generation~\cite{hsiao2020single}.
Integration with two-dimensional electronic systems has been used to investigate high-frequency acoustically driven transport and band-structure engineering ~\cite{bandhu2016controlling, hernandez2018interaction, Lane2018, byeon2021piezoacoustics, Zhao2022, Fang2023, Meril2025acoustoelectricsuperlattices}.
 
Beyond condensed matter physics, SAW devices operating in the MHz to GHz frequency range are ubiquitous in RF and microwave signal processing, functioning as filters, delay lines, and resonators~\cite{Morgan}. 
Moreover, since SAWs propagate on the surface of the substrate, they are extremely sensitive to minute perturbations of the crystal surface due to external factors such as pressure, temperature, humidity, mass loading, or the presence of specific chemical species. 
This inherent surface selective sensitivity makes SAW devices exceptionally effective as sensors for environmental monitoring~\cite{memon2022surface, galipeau1991study,wang2015surface, he2013high}, biosensing~\cite{huang2021surface,baumgartner2023recent,lange2008surface,fourati2023applications,pohanka2018overview,mujahid2019overview,hartz2020lateral,naranda2022practical,baumgartner2023recent,skladal2024piezoelectric}, and chemical sensing~\cite{gomes2001application,kuchmenko2019perspective,li2023advances}. 

Recent advances in surface acoustic wave (SAW) devices have also drawn significant attention in quantum information science, where solid-state platforms such as superconducting qubits~\cite{gustafsson2014propagating, manenti2017circuit, Satzinger2018, Moores2018, kitzman2023phononic}, semiconductor quantum dots~\cite{mcneil2011demand, gell2008modulation, kataoka2007single, patel2024surface, wang2024gated}, and defect centers~\cite{lee2017topical, whiteley2019spin, PhysRevX.5.031031}, are coupled to surface phononic modes. These hybrid systems have enabled mechanically mediated quantum information processing experiments~\cite{qiao2023splitting, dumur2021quantum, hann2019hardware, bienfait2019phonon,dumur2021quantum}, phonon-assisted transport of quantum states over mesoscopic distances~\cite{hermelin2011electrons, mcneil2011demand,bertrand2016fast,wiele1998photon,couto2009photon}, and microwave frequency optomechanical sideband generation ~\cite{golter2016coupling,metcalfe2010resolved}, while also offering prospects for microwave-to-optical transduction~\cite{Shumeiko2016}. 

In addition to quantum information processing, SAW devices can also function as quantum-assisted sensors, capable of detecting excitations within the coupled qubit-phonon system~\cite{kitzman2023quantum, cleland2024studying}. 
Since SAW devices are well-established sensors in both classical and quantum applications, an enhancement to their sensitivity could significantly impact an extremely wide range of disciplines. 

The performance of SAW resonator-based sensors can be improved by increasing their sensitivity to external stimuli, typically observed as shifts in the resonance frequency of the confined piezoelectric mode, and by improving the precision and stability of center frequency measurements~\cite{rath2025frequency}. While advances in piezoelectric materials~\cite{ding2020enhanced,nicolay2018ln,muller2017gan} and device structures~\cite{wang2015diaphragm,li2020optimization} can boost sensitivity, precise measurement of SAW frequency shifts can be further improved by engineering the spectral response of the device. In particular, the systematic introduction of interference effects into the resonator can shape the SAW spectrum such that even small frequency shifts result in pronounced piezoacoustic amplitude variations, enhancing the detectability of frequency shifts produced by environmental changes. 
Fano interference~\cite{Fano1961} has emerged as a promising mechanism for increasing the sensing capabilities of mechanical and optical resonator-based systems~\cite{wu2012fano, qu2023research, pathania2021fano, limonov2021fano, liu2012multiple, mohamed2024fano}. This type of interference occurs when there is interaction between a resonant mode and a continuum background in a resonator structure. 
This interference results in a characteristic asymmetry in the shape of the composite spectral line of the resonator, which is parameterized by the so-called Fano factor $q$, as shown in Figure~1~(a). Physically, $q$ encodes the relative strength and phase between the resonant signal pathway and the continuum background. A larger value of $|q|$ signifies a stronger contribution from the resonant channel, resulting in a more pronounced spectral asymmetry. This spectral skew can lead to significant amplitude variations for small changes in the center frequency of the resonator, thus improving sensitivity to external perturbations. While increasing the intrinsic quality factor of a resonator also improves frequency resolution, Fano interference introduces an additional method for tailoring the resonator spectrum, enabling improved transduction of small perturbations into measurable changes in the resonator transmission, even when the intrinsic quality factor is moderate. Previous studies have reported the existence of Fano interference in SAW–based hybrid quantum acoustic systems containing qubits~\cite{kitzman2023quantum, sarabalis2020s}, and in classical SAW devices through the introduction of externally connected shunt capacitance~\cite{wang2024putting}. 

\begin{figure}[ht]
    \centering\includegraphics[width=1\linewidth]{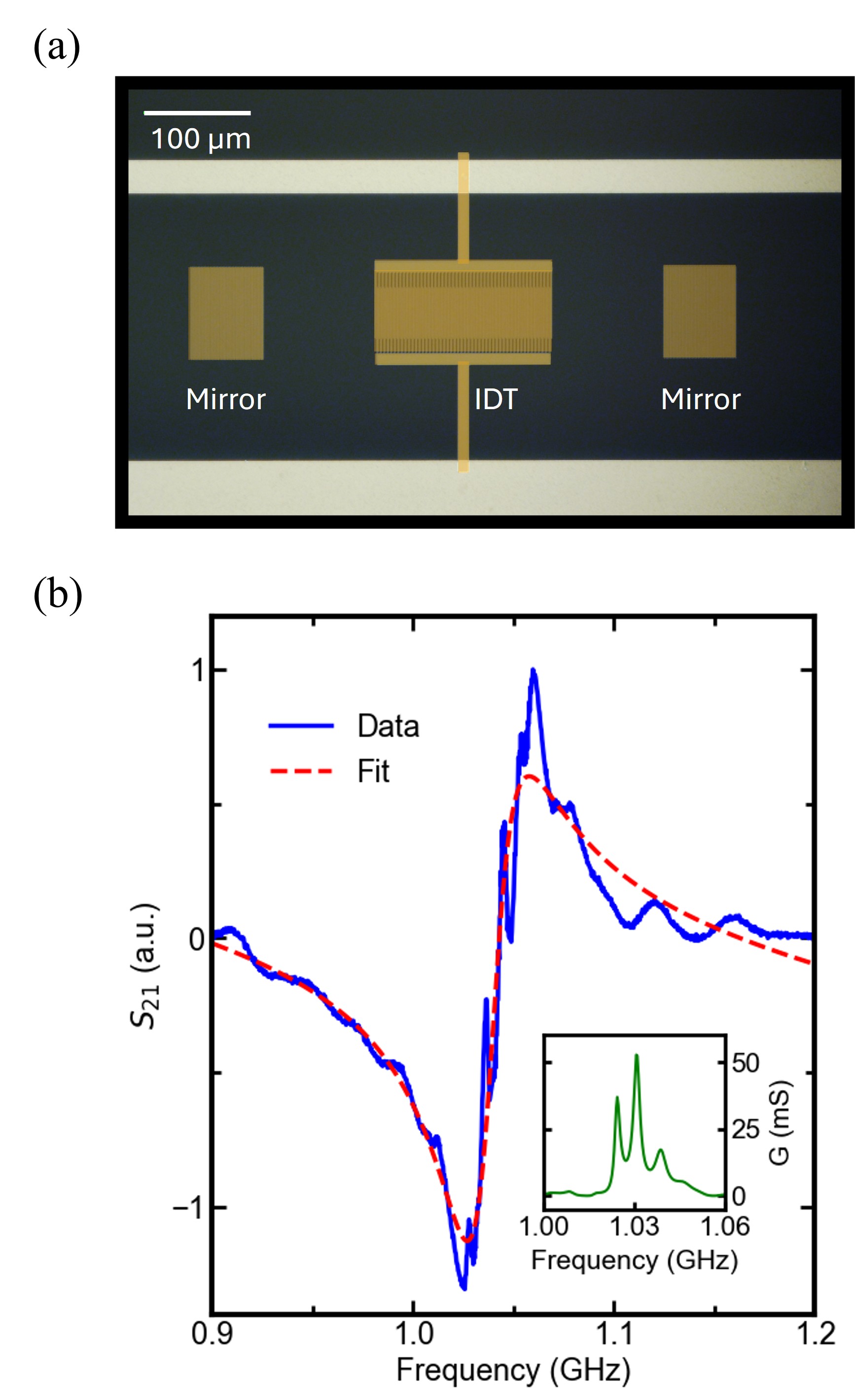}
    \caption{~(a) Optical false colored image of the SAW resonator (Device~\uppercase\expandafter{\romannumeral 1}), which consists of a central IDT structure, and acoustic Bragg mirrors on both sides of the IDT. The IDT is connected galvanically on one side to the center pin of a coplanar waveguide while the other side of the IDT is attached to ground.~(b) Transmission spectrum $S_{21}$ (blue data) of the SAW device showing the expected discrete resonant acoustic modes as dips, along with a characteristic asymmetric line shape corresponding to composite Fano interference in the device. The red dashed curve corresponds to a fit using Eq.~1. The expected response of the SAW resonator was designed using the coupling of modes method~\cite{Morgan} as shown in the inset, which presents the effective electrical conductance $G$ of the multi-mode SAW resonator.}  
\end{figure}

In this work, we employ a different approach to systematically, and simultaneously, introduce both microwave frequency electromagnetic and acoustic Fano interference into SAW resonators. We achieve this by leveraging an established notch-type microwave control and readout system to produce GHz-frequency electromagnetic interference~\cite{Rieger2023} while tailoring the acoustic resonator design to support multiple discrete SAW modes and continuum background surface phonon states. We systematically modify the acoustic reflectivity of the resonator mirrors to independently identify and distinguish the contributions of the electromagnetic and acoustic interference channels, as well as their impact on the device’s spectral response. We also investigate the evolution of the composite Fano interference as the device is cooled down to $T \simeq 1$~K from room temperature.

\section{Experimental setup}

\begin{figure*}[ht]
    \centering
    \includegraphics[width=0.86\linewidth]{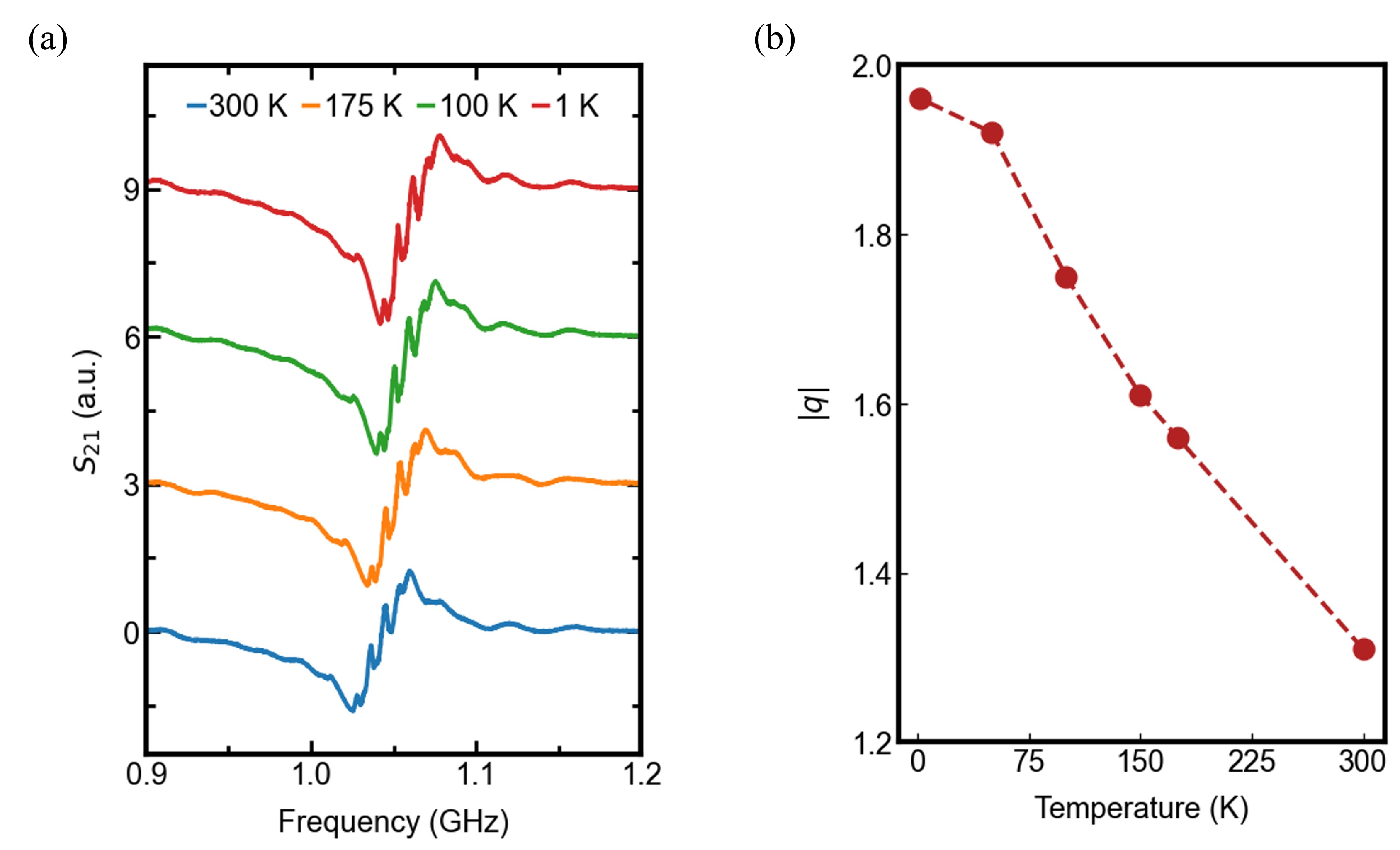}
    \caption{~(a) The spectral response of the SAW resonator (Device~\uppercase\expandafter{\romannumeral 1}) at different temperatures. The curves have been vertically shifted relative to each other for clarity. As the temperature decreases, the resonant acoustic modes shift to higher frequencies and the asymmetry of the overall line shape increases.~(b) Dependence of the magnitude of the Fano parameter $|q|$ versus temperature. With decreasing temperature, $|q|$ increases, indicating an enhancement of the resonant acoustic pathway over the non-resonant electromagnetic background channel.}
\end{figure*}

The experimental setup, schematically illustrated in Figure~1~(b), employs a notch-type transmission measurement configuration of a SAW resonator connected to a microwave waveguide line for excitation and measurement. 
This configuration was specifically adopted to systematically introduce both acoustic and electromagnetic Fano interference into the surface acoustic wave spectrum. 
To independently isolate and study the contributions of these two types of interference, three distinct devices (Devices~\uppercase\expandafter{\romannumeral 1}, ~\uppercase\expandafter{\romannumeral 2} and~\uppercase\expandafter{\romannumeral 3}) featuring SAW resonators with differing levels of acoustic confinement were fabricated and investigated. 

Each device is a SAW resonator consisting of a central interdigitated transducer (IDT) having 50 pairs of fingers, as shown in Figure~2~(a). When a microwave signal is applied across the IDT, the resulting electric field induces periodic stress and strain via the piezoelectric effect, generating surface acoustic waves that propagate along the substrate. Devices~\uppercase\expandafter{\romannumeral 1} and~\uppercase\expandafter{\romannumeral 2} include Bragg mirrors on either side of the IDT, composed of 20 and 10 finger pairs per mirror, respectively, which are separated from the IDT by a free propagation distance of $L_{\textrm{free}}=110~\mu$m. 
These mirrors provide periodic impedance modulation that reflects the SAWs back toward the IDT, confining the surface waves between the two reflector arrays. This reflection-induced confinement forms a resonant cavity, enhancing the acoustic amplitude and produces well-defined resonances (see Figure 2~(b)). In contrast, Device~\uppercase\expandafter{\romannumeral 3} does not have Bragg mirrors and serves as a minimally confined reference device having an IDT response only.
All components—including the IDT, Bragg mirrors, transmission line, and ground plane were fabricated on a Y-cut LiNbO$_3$ piezoelectric substrate. The structures were patterned using a combination of photolithography and electron-beam lithography, and the aluminum layers were subsequently deposited via thermal evaporation. The device geometries were designed using coupling-of-modes simulations to generate surface acoustic waves having a wavelength of $3.44~\mu$m, supporting multimode resonances centered around 1.040~GHz and a broader continuum background, as shown in the inset of Figure 2~(b). In each device, the SAW resonator is galvanically connected to a coplanar transmission line waveguide. One side of the resonator connects to the central microwave line of the waveguide, while the other connects to an outer ground plane (see Fig.~1~(b) and Fig.~2~(a)). This configuration enables both efficient electrical excitation of SAWs and piezoelectric back-conversion of acoustic waves into measurable electrical signals. Importantly, this configuration also introduces a channel for microwave frequency electromagnetic Fano interference into the composite system~\cite{Rieger2023}, as will be discussed in detail in later sections.

\section{Results and discussion}
The measured spectral response of Device~\uppercase\expandafter{\romannumeral 1} is shown in {Figure 2~(b)}. As expected, the spectrum exhibits the multiple discrete resonant SAW modes, which appear as dips in the central region of the frequency spectrum, along with a distinctly spectral asymmetry envelope characteristic of Fano interference. 
The spectral response of a Fano resonance is commonly modeled with the following lineshape function,
\begin{equation}
   f (\omega) = A\left [ \frac {\left ( q \frac{\gamma} {2} + \omega - \omega_0 \right)^2} {\left( \frac{\gamma}{2}\right)^2 + (\omega - \omega_0)^2}\right],
\end{equation}
where $\omega$ and $\omega_0$ denote the applied frequency and the resonator center frequency, respectively, and $\gamma$ corresponds to the full width at half maximum of the resonance spectrum. 
In effect, Eq.~1 is a modified Lorentzian in which the numerator accounts for the coherent superposition of resonant and non-resonant scattering pathways. 
The dimensionless Fano parameter $q$ quantifies the relative amplitude and phase of these two pathways and thus controls the degree of spectral asymmetry.
The scaling factor $A$ sets the overall magnitude of the spectral response, while the detuning $\omega - \omega_0$ defines the frequency offset from the center resonance.

We fit the asymmetric SAW resonator spectrum using Eq.~1 with the results shown in Figure~3~(b) and find that it captures the overall Fano lineshape of the envelope of the spectrum. As we have noted previously, the Fano interference in this device arises from both acoustic and electromagnetic contributions. The acoustic contribution originates from the interference between the SAW resonant modes and the continuum acoustic background~\cite{kitzman2023quantum}. However, the electromagnetic contribution stems from the notch-type measurement configuration employed, in which the scattered microwave signal from the SAW resonator interferes with the background signal transmitted through the transmission line waveguide~\cite{Rieger2023}. This composite Fano interference is illustrated in Figure 1~(b). When the electromagnetic signal introduced through the transmission line is in resonance with the acoustic resonator, it is transduced into both resonant and continuum SAW excitations, which interfere with each other, giving rise to acoustic Fano interference. This acoustic signal is then transduced back into an electromagnetic signal, which subsequently interferes with the background microwave transmission through the waveguide, which introduces a purely electromagnetic component of Fano interference.

To begin to understand the interplay between these two interference channels we investigated their temperature dependence by cooling the SAW devices down to $T\simeq 1$~K in a pumped helium cryostat. 
The transmission spectra recorded at multiple temperatures are presented in Figure 3~(a) and show a clear and systematic blueshift in resonance frequency with decreasing temperature. This increase in frequency can be attributed to an enhancement in the acoustic phase velocity of the resonator upon cooling, since the resonance frequency is given by $f_0 = v_{SAW}/\lambda$, where $f_0$, $v_{SAW}$ and $\lambda$ are the central resonant frequency, phase velocity, and acoustic wavelength (set lithographically by the IDT and mirror spacing) of the SAW resonator, respectively. The dominant mechanism responsible for the increase in SAW velocity with decreasing temperature is the increase in the elastic stiffness of the substrate as bulk phonons are progressively suppressed. 
Additionally, the SAW velocity also depends on the piezoelectric properties of the substrate hosting the SAWs, specifically
\begin{equation}
    v_{\text{SAW}} = v_0 \left(1 + \frac{K^2}{2}\right),
\end{equation}
where $v_0$ is the SAW velocity produced by purely elastic properties of the substrate, and $K$ is the electromechanical coupling coefficient of the substrate, which parameterizes the strength of the piezoelectricity. A small but systematic reduction in the dielectric constant of LiNbO$_3$ with decreasing temperature~\cite{mansingh1985ac} leads to an increase in the piezoelectric coupling constant, since $K^2~\propto~{e_{ijk}^2}/{\varepsilon^{T} s_{ijk}}$,
where $e_{ijk}$ is the piezoelectric stress coefficient tensor, $\varepsilon_{T}$ dielectric constant at constant stress, and $s_{ijk}$ is the elastic stiffness coefficient tensor of the substrate~\cite{Morgan}. This increase in $K$ leads to a corresponding increase in $v_{\text{SAW}}$ upon cooling the substrate. We note that this mechanism is relatively more significant in strongly piezoelectric materials such as LiNbO$_3$, for which $K^2 \sim 4\%\text{-}6\%$~\cite{Morgan}.

Beyond the frequency shift, we also observe a notable evolution in the spectral asymmetry upon cooling, indicating that the Fano interference mechanisms present in the device are sensitive to temperature changes. To quantify this behavior, we fit the spectral line shapes at different temperatures using Eq.~1, and in Figure~3~(b) we plot the corresponding Fano parameter as a function of temperature. We find that the magnitude of the Fano parameter $|q|$ increases with decreasing temperature. In general, the Fano parameter quantifies the relative weight of the resonant and continuum contributions to the overall spectral lineshape~\cite{iizawa2021quantum}. 
Therefore, the increase in $|q|$ that we observe suggests an enhancement of the resonant electromagnetic pathway over the non-resonant background channels. This could be due to the reduction of the dielectric constant at lower temperatures, enhancing the electromechanical coupling in the piezoelectric material, and strengthening the transduction of the resonant acoustic signals.

\begin{figure}[ht]
    \centering
    \includegraphics[width=0.9\linewidth]{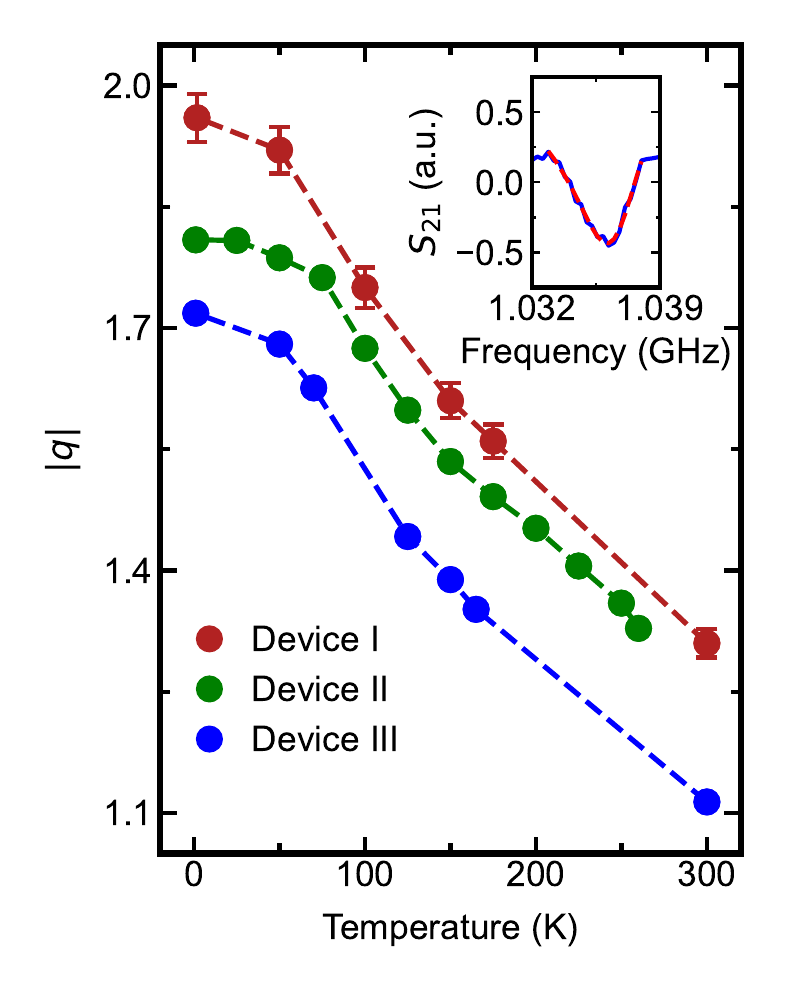}
    \caption{Fano parameter as a function of temperature for the three SAW Devices (Devices~\uppercase\expandafter{\romannumeral 1}, ~\uppercase\expandafter{\romannumeral 2}, and ~\uppercase\expandafter{\romannumeral3}) with varying acoustic confinement. Device~\uppercase\expandafter{\romannumeral 1}, having 20 pairs of Bragg mirror elements, (red points) exhibits the highest value of $|q|$ across the entire temperature range. As the surface acoustic wave confinement is reduced in the other devices (i.e.Device~\uppercase\expandafter{\romannumeral 2} with 10 pairs of mirror elements (green points), and   Device~\uppercase\expandafter{\romannumeral 3} with no mirrors (blue points)) we observe a reduction in $|q|$, indicative of a systematic reduction of phononic interference effects with decreasing acoustic confinement. Inset: Magnitude spectrum of an individual confined acoustic mode from Device~I (blue) and corresponding fit using Eq.~1 (red dashed curve) with $|q|=0.28$.}
\end{figure}

Although both electromagnetic and acoustic contributions to Fano interference are present in our device, and collectively enhance its sensitivity to frequency shifts, unambiguously disentangling these contributions is crucial for a clear understanding of the underlying interference mechanisms and for ultimately controlling them to shape the SAW resonator spectrum. This requires selective modification of one contribution to isolate the other.
The acoustic contribution can be systematically enhanced or suppressed by adjusting the reflectivity of acoustic Bragg mirrors confining the SAW resonances. We performed spectroscopic measurements over a broad range of temperature on Devices~\uppercase\expandafter{\romannumeral 2} and Devices~\uppercase\expandafter{\romannumeral 3}, which were designed to have reduced acoustic reflectivity by decreasing the number of finger pairs in the Bragg mirrors or by completely removing the mirrors altogether. As expected, Device~\uppercase\expandafter{\romannumeral 3}, having no acoustic mirrors, shows no trace of discrete resonant modes in its spectral response. Nevertheless, it exhibits a characteristic asymmetric Fano lineshape. The absence of mirrors precludes the formation of sharp standing-wave acoustic cavity modes; however, the IDT itself is a frequency-selective transducer, generating SAWs over a relatively broad resonance. This provides a frequency-selective acoustic transduction pathway for the microwave signal, which interferes with the direct, and comparatively much broader, continuum microwave background propagating through the transmission line. This results in the observed Fano lineshape in Device~\uppercase\expandafter{\romannumeral 3}, which is predominately electromagnetic in nature, and a markedly smaller Fano parameter $|q|$ for Device~\uppercase\expandafter{\romannumeral 3} (see Fig.~4, blue points), compared to Device~\uppercase\expandafter{\romannumeral 1} (see Fig.~4, red points). Consistent with this interpretation, the values of $|q|$ obtained for Device~\uppercase\expandafter{\romannumeral 2} (see Fig.~4, green points) lie between those of Devices~\uppercase\expandafter{\romannumeral 1} and Devices~\uppercase\expandafter{\romannumeral 3}, highlighting the systematic reduction of acoustic interference effects with decreasing SAW confinement. These observations confirm the coexistence of acoustic and electromagnetic Fano interference in Device~\uppercase\expandafter{\romannumeral 1} and demonstrate that Bragg mirror design can be used as a tuning knob to vary the spectral symmetry of SAW resonators. 
We note that the temperature dependence of the Fano parameter across all three devices exhibits a consistent trend, indicating an enhancement of the resonant SAW modes at lower temperatures, in agreement with the discussion presented earlier for Devices~\uppercase\expandafter{\romannumeral 1}. 

Additionally, a closer examination of Device~I, where the confined acoustic modes are well-resolved, reveals Fano interference associated with the individual acoustic resonances themselves. These mode-specific asymmetries originate predominantly from acoustic interference within the SAW resonator. The extracted Fano parameters for the individual acoustic modes range from approximately 0.18 to 0.28, consistent with previous observations~\cite{kitzman2023quantum}. A representative resonance and its corresponding Fano fit are shown in the inset of Figure~4, for which $|q|=0.28$.

\section{Conclusion}
In summary, we have experimentally investigated the introduction of Fano interference into SAW-based resonator devices. The notch-type measurement configuration leads to the emergence of a Fano resonance, arising from both acoustic and electromagnetic interference. We systematically modify the acoustic confinement across several devices to disentangle the role of each contribution to the composite Fano interference. By analyzing the measured spectral data over a wide range of temperatures, we extract the temperature-dependent Fano factor $|q(T)|$, which parameterizes the strength of the interference effects in the device and highlights the potential of this type of SAW-based resonator, for variable-temperature sensing. 

Beyond demonstrating an effective strategy to enhance the sensitivity of SAW-based devices, our results open several promising directions for future research. Developing approaches to selectively suppress or enhance the electromagnetic Fano interference could enable fully tunable sensor platforms, in which sensitivity can be independently controlled through the electromagnetic and acoustic channels. Furthermore, the persistence of Fano interference at low temperatures suggests that these engineered SAW resonators could be integrated with superconducting qubits, paving the way toward next-generation qubit-assisted quantum acoustic sensors.

\begin{acknowledgments}
We would like to thank K.W.~Murch and A.J.~Schleusner for valuable discussions. We also thank R.~Loloee and B.~Bi for technical assistance and use of the W.~M.~Keck Microfabrication Facility at Michigan State University. This work was supported by NSF Grant No. ECCS-2142846 (CAREER) and a Targeted Support Grant for Technology Development (TSGTD) from MSU. Additionally JP acknowledges support from the Cowen Family Endowment at MSU and from the Gordon and Betty Moore Foundation under Grant DOI~10.37807/GBMF13719
\end{acknowledgments}

\end{document}